\documentclass[%
reprint,
superscriptaddress,
 amsmath,amssymb,
 aps,
]{revtex4-2}

\usepackage{graphicx}% Include figure files
\usepackage{dcolumn}% Align table columns on decimal point
\usepackage{bm}% bold math
\usepackage{color}
\usepackage{cancel}
\usepackage{comment}
\usepackage{appendix}
\usepackage{braket}
\allowdisplaybreaks
\usepackage[colorlinks=true,
  allcolors={blue}]{hyperref}% add hypertext capabilities
\usepackage{cleveref}
\begin{document}

%\title{Conventional superconductivity in tetragonal C$_3$N$_4$ with hole doping}

\title{Strong electron-phonon coupling and 
phonon-induced superconductivity \\in tetragonal C$_3$N$_4$ with hole doping}

\author{Alexander N. Rudenko}
\email{a.rudenko@science.ru.nl}

\author{Danis I. Badrtdinov}
\affiliation{\mbox{Radboud University, Institute for Molecules and Materials, Heijendaalseweg 135, 6525AJ Nijmegen, The Netherlands}}

\author{Igor A. Abrikosov}
\affiliation{\mbox{Department of Physics, Chemistry, and Biology (IFM),
Link\"{o}ping University, SE-581 83, Link\"oping, Sweden}}

\author{Mikhail I. Katsnelson}
\affiliation{\mbox{Radboud University, Institute for Molecules and Materials, Heijendaalseweg 135, 6525AJ Nijmegen, The Netherlands}}

\date{\today}% It is always \today, today,

\begin{abstract}
	C$_3$N$_4$ is a recently discovered phase of carbon nitrides with the tetragonal crystal structure [D.~Laniel \emph{et al.}, \href{https://doi.org/10.1002/adma.202308030}{Adv.~Mater. 2023, 2308030}] that is stable at ambient conditions. C$_3$N$_4$ is a
%n air-stable 
semiconductor exhibiting flat-band anomalies in the valence band, suggesting the emergence of many-body instabilities upon hole doping. Here, using state-of-the-art first-principles calculations we show that hole-doped C$_3$N$_4$ reveals strong electron-phonon coupling, leading to the formation of a gapped superconducting state. The phase transition temperatures turn out to be strongly dependent on the hole concentration. We propose that holes could be injected into C$_3$N$_4$
via boron doping which induces, according to our 
%computational 
results, a rigid shift of the Fermi energy without significant modification of the electronic structure. Based on 
%ab initio 
the electron-phonon coupling and Coulomb pseudopotential calculated from first principles, we conclude that the boron concentration of 6 atoms per nm$^3$ would be required to reach the critical temperature of $\sim$36~K at ambient pressure.
\end{abstract}

\maketitle

%\tableofcontents

\section{Introduction}
The prospects of high-temperature superconductivity attract enormous attention to this phenomenon \cite{FloresLivas2020,Lilia2022,Pickett2023}. In this respect, the performance of conventional superconductors, i.e., those mediated by the electron-phonon pairing mechanism, are believed to be limited. Nevertheless, there has been considerable progress in the experimental discovery of phonon-mediated superconductivity in hydrides with reported critical temperatures $T_\mathrm{c}$ of up to 260~K at extremely high pressures (above 100 GPa) \cite{Drozdov2015,Hemley2019,Drozdov2019,Sneider2021,Kong2021}. A promising direction in attaining high $T_\mathrm{c}$ is to focus on compounds with light elements that permit the existence of high-frequency phonons strongly coupled to charge carriers. This strategy was suggested by Ashcroft \cite{Ashcroft} as a practical realization of his old idea of superconductivity of metallic hydrogen \cite{Ashcroft_old}. From this perspective, the role of heavy elements is just to provide a container for solid hydrogen, assuming that the hydrogen contribution to the electron states at the Fermi level is essential.

Other light elements are obvious candidates for obtaining superconductors with reasonably high $T_\mathrm{c}$. MgB$_2$ with $T_\mathrm{c} \approx 40$~K is probably the most known example \cite{Nagamatsu2001,Mazin2001}. As for the next to boron element in the periodic table, carbon, intercalated graphite and graphene laminates were also intensively studied both experimentally \cite{Weller2005,Heguri2015,Chapman2016} and theoretically \cite{Mauri,Margine2014,Margine2016,Zheng2016} but the reached $T_\mathrm{c}$'s are always essentially smaller than for MgB$_2$. Here, we predict theoretically the phonon-mediated superconductivity in another group of light-element compounds, namely, carbon nitrides, with the highest critical temperatures of $\sim 36$~K at ambient pressure.

A number of previously unknown crystalline phases of nitrogen and its compounds have been discovered in the past years by means of high-pressure synthesis \cite{Eremets2004,Laniel2020,Bykov2021}. Recently, the synthesis of ultra-incompressible carbon nitrides recoverable at ambient conditions has been reported \cite{Laniel2023}. Among them, a tetragonal phase C$_3$N$_4$ with the space group $I\bar{4}2m$ demonstrates remarkable electronic properties. Particularly, C$_3$N$_4$ is a semiconductor with the flat-band features near the top of the valence band, suggesting the essential role of many-body effects upon hole doping. In combination with high-frequency 
phonons with the highest energy of $\sim$160 meV, such characteristics make C$_3$N$_4$ an appealing candidate for superconductivity, which constitutes the main focus of our study.

\section{Computational details}
First-principles electronic structure and total energy calculations were performed using density functional theory (DFT) within the plane-wave pseudopotential method \cite{Vanderbilt1990} as implemented in the {\sc quantum espresso} distribution \cite{Giannozzi2009,Giannozzi2017}.
The exchange-correlation effects were considered within the generalized-gradient approximation (GGA) functional in the Perdew-Burke-Ernzerhof parametrization \cite{pbe}. 
Ultrasoft pseudopotentials were used with a 50 Ry energy cutoff for the plane waves and 400 Ry for the charge density \cite{pseudo}. A convergence threshold for the total energy of $10^{-12}$ Ry was used in self-consistent calculations. The atomic structure was relaxed until the residual force components of each atom were less than  10$^{-4}$ Ry$\cdot$\AA$^{-1}$.
The lattice dynamics and phonon-related properties were calculated within the density-functional perturbation theory (DFPT) \cite{Baroni2001}.
To achieve convergence in the calculation of the superconducting properties, the electron-phonon interactions were interpolated using the Wannier functions \cite{Giustino2007,Giustino2017} as implemented in the {\sc epw} package \cite{EPW}. The Wannier functions were constructed from the manifold of four valence states using the procedure of maximum localization \cite{Marzari2012,wannier90}, resulting in one $p$-like orbital per N atom.
The Brillouin zone was sampled by (12$\times$12$\times$12) ${\bf k}$-point and (6$\times$6$\times$6) ${\bf q}$-point meshes in the DFT and DFPT calculations, respectively. 
The Brillouin zone (BZ) integrals were calculated using interpolated ${\bf k}$ and ${\bf q}$ meshes.
The numerical integration was performed using the Gaussian smearing with the parameters 10 and 0.5 meV for electrons and phonons, respectively. Convergence tests of the BZ integrals are provided in the Supplemental Material (SM) \cite{SM}.

\begin{figure}[tb]
\centering
\includegraphics[width  =1.00\linewidth]{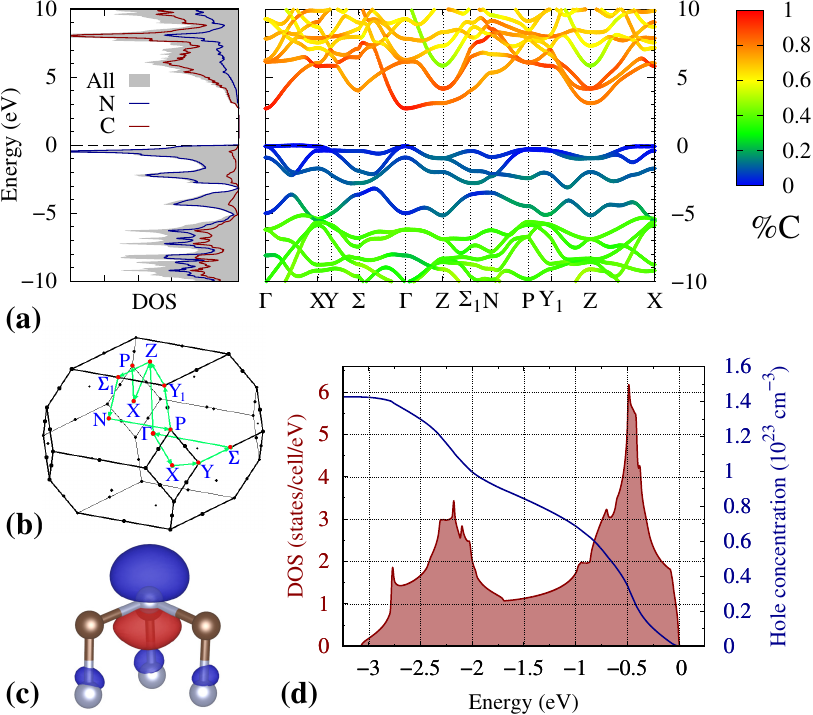}
\caption{
	(a) DOS and band structure projected onto the $sp$ states of N and C atoms in C$_3$N$_4$. Color indicates the contribution of N (blue) or C (red) states. (b) The Brillouin zone of the tetragonal lattice of C$_3$N$_4$ with the high-symmetry directions along which the band structure is calculated. (c) $p_z$-like Wannier orbital residing on N atoms describing the valence states in C$_3$N$_4$. (d) DOS of the valence band (shaded area) and the hole concentration (blue solid line) shown as a function of energy. Zero energy corresponds to the valence band top.}
\label{bands}
\end{figure}

Dielectric screening and the Coulomb interactions in undoped C$_3$N$_4$ were calculated in the Wannier basis using the random phase approximation (RPA) as implemented in {\sc vasp} \cite{vasp-gw1,vasp-gw2,KaltakcRPA}.
To this end, we carried out calculations within the projector augmented wave formalism (PAW) \cite{paw2} using GGA and a 400 eV cutoff for the plane waves. In order to rule out possible inconsistencies between {\sc vasp} and {\sc quantum espresso}, we performed a comparison between the band structures calculated using these codes, which are found to be almost identical \cite{SM}.
The polarization function for RPA calculations was calculated taking a window of $\sim$100 eV for the empty states.
The calculation of the Coulomb interactions and screening in doped C$_3$N$_4$ was performed using an in-house developed code.

\section{Results}
\begin{figure}[!tbp]
\centering
\includegraphics[width=0.9\linewidth]{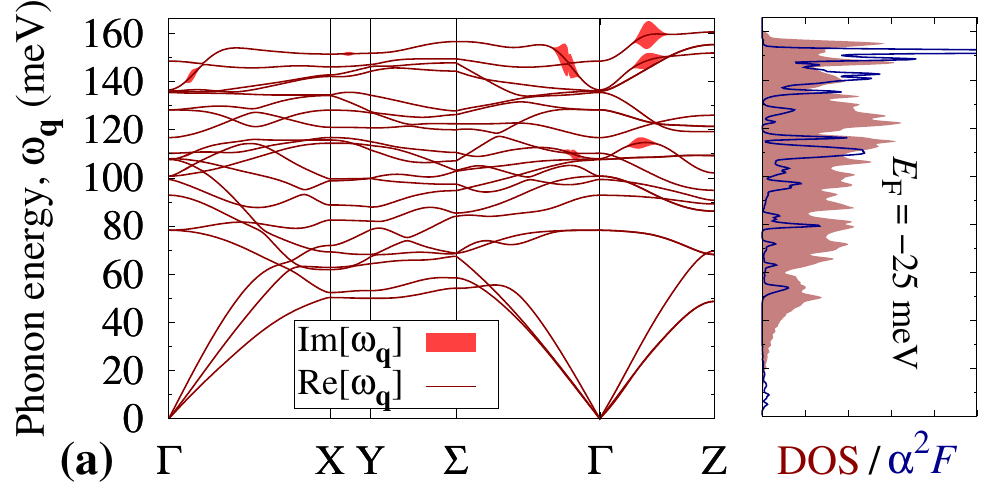}
\mbox{
\includegraphics[width=0.50\linewidth]{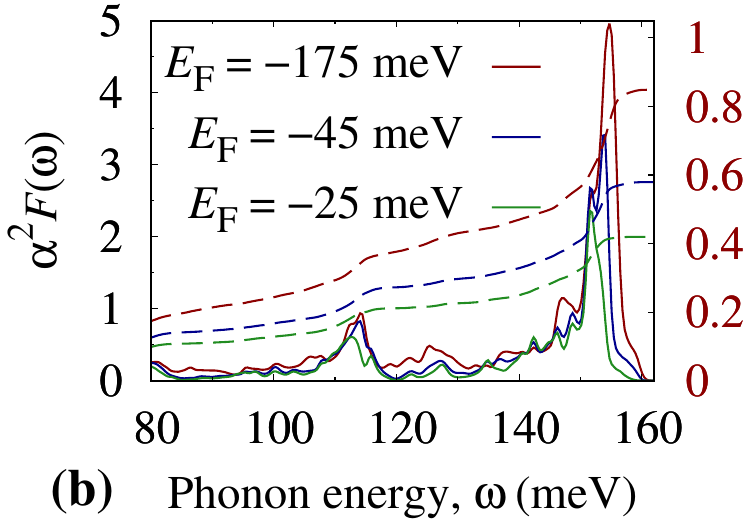}
\includegraphics[width=0.50\linewidth]{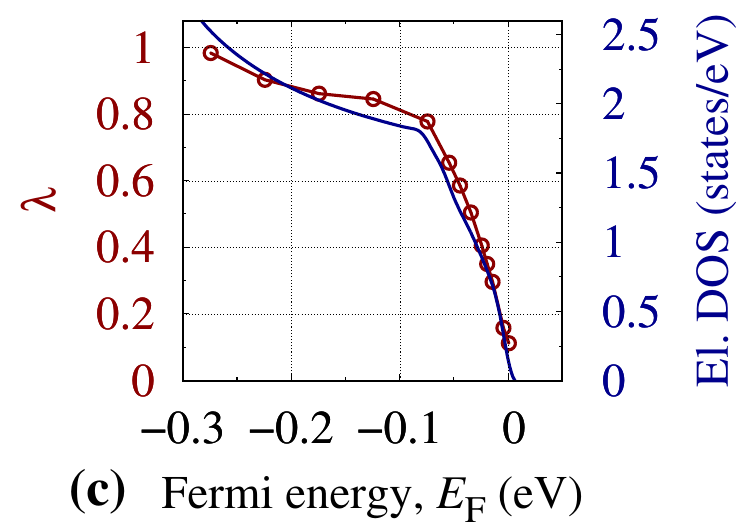}
}
\caption{
(a) Phonon dispersion (lines) and the corresponding linewidths (filled areas) calculated for the Fermi energy $E_\mathrm{F}=-25$~meV and $T=50$~K along selected high-symmetry directions of the Brillouin zone in C$_3$N$_4$. Right panel shows phonon DOS (shaded area) and the Eliashberg function $\alpha^2F(\omega)$ (blue solid line) calculated for $E_\mathrm{F}=-25$~meV. (b) High-frequency region of $\alpha^2F(\omega)$ plotted for three different $E_\mathrm{F}$. Dashed line shows $\lambda$. (c) Electron-phonon constant $\lambda$ shown as a function of the Fermi energy $E_\mathrm{F}$ superimposed onto the electron DOS. 
}
\label{phonons}
\end{figure}

\subsection{Electronic structure}
Figure~\ref{bands}(a) shows the band structure and density of states (DOS) calculated for C$_3$N$_4$ along the high-symmetry directions of the Brillouin zone shown in Fig.~\ref{bands}(b). From Fig.~\ref{bands}(a) one can see that the valence band is predominantly formed by the $sp$ states of N atoms, while C atoms mainly contribute to the conduction band.  The valence N states are well isolated within the range from $-$5 to 0 eV. The subspace of N valence states can be represented in terms of equivalent $p_z$-like Wannier orbitals, each centered on N atoms, as visualized in Fig.~\ref{bands}(c). The GGA electronic structure of C$_3$N$_4$ exhibits an indirect band gap of $\sim2.7$ eV. The conduction band bottom is located at the $\Gamma$ point, whereas the valence band top is in between the $\Gamma$ and $X$ points, forming a saddle point at the $\Gamma$ point $\sim$0.06 eV below the valence band top. This saddle point gives rise to a van Hove singularity (vHS) which, however, does not form an isolated peak in DOS. At the same time, there is another vHS close to the valence band top at $\sim-$0.5 eV, originating from the flat band along the Y--$\Sigma$ direction, giving rise to a prominent peak in DOS. The proximity of the vHS to the valence band edge suggests that the system could be potentially $p$-doped in order to align the vHS with the Fermi energy. This, in turn, might lead to the emergence of many-body effects and exotic physics. Practically, hole doping in C$_3$N$_4$ could be achieved by the doping of B atoms, which tend to replace C atoms, which is discussed below. The dependence of the hole concentration on the Fermi energy is shown in Fig.~\ref{bands}(d). The Fermi surfaces for relevant doping levels are presented in the SM \cite{SM}.

\subsection{Phonons and electron-phonon coupling}

Figure \ref{phonons}(a) shows the phonon dispersion $\omega_{\bf q\nu}$ along with the phonon linewidths 
$\gamma_{\bf q\nu}\sim N_\mathrm{F}\lambda_{\bf q\nu}\omega^2_{\bf q\nu}$ 
calculated at $T=50$~K and $E_\mathrm{F}=-25$~meV from the imaginary part of the phonon self-energy \cite{Giustino2017}. The phonon spectrum exhibits three linearly dispersing  branches in the range up to $\sim$50 meV and a bunch of optical branches extending up to 160 meV, similar to diamond \cite{PhysRevB.48.3156}. The high-frequency character of optical phonons in C$_3$N$_4$ is determined by the low atomic masses and strong interatomic bonding. Overall, the spectrum is typical for a 3D compound, showing no anomalies. At high frequencies, there is an apparent broadening of the phonon lines, i.e., along the $\Sigma$--$\Gamma$ and $\Gamma$--$Z$ directions, indicating strong electron-phonon interaction at specific wave vectors ${\bf q}$. This gives rise to a sharp high-energy peak in the Eliashberg function \begin{equation}
    \alpha^2F(\omega) = \frac{1}{2}\sum_{{\bf q}\nu} \omega_{{\bf q}\nu}\lambda_{{\bf q}\nu}\delta(\omega - \omega_{{\bf q}\nu}),
    \label{a2f}
\end{equation}
shown 
in Fig.~\ref{phonons}(b) for three different $E_\mathrm{F}$, from which one can see that approximately half of the contribution to the electron-phonon coupling originates from the high-energy region. In Eq.~(\ref{a2f}), $\lambda_{{\bf q}\nu}=\langle |g^{\nu}_{\bf q}|^2 \rangle / \omega_{{\bf q}\nu}$ is the momentum-resolved electron-phonon coupling constant averaged over the Fermi surface with
\begin{equation}
g^{nn',\nu}_{{\bf k}{\bf k}'}=\sqrt{\frac{\hbar}{2m_0\omega_{{\bf q}\nu}}}\langle \psi_{n'{\bf k}'}| \partial_{-{\bf q}\nu}V |\psi_{n{\bf k}} \rangle 
\end{equation}
being the matrix element of the electron-phonon interaction, where $\partial_{-{\bf q}\nu}V$ is the phonon-induced variation of the electronic potential written in the basis of Bloch states $\psi_{n{\bf k}}$. 
Figure~\ref{phonons}(c) shows the integrated electron-phonon coupling $\lambda = 2\int d\omega \frac{\alpha^2F(\omega)}{\omega}$ as a function of $E_\mathrm{F}$. Already at $E_\mathrm{F}=-50$~meV, we obtain $\lambda = 0.6$, which can be qualified as a strong regime \cite{Allen1983}. The energy dependence of $\lambda$ closely follows electronic DOS, indicating that the strong coupling in C$_3$N$_4$ is governed by the peculiarities of its electronic structure. 

At the same time, in the absence of magnetism, large $\lambda$ suggests an instability with respect to the formation of a superconducting state. Ignoring the Coulomb interaction, a simple estimate using the Allen-Dynes formula \cite{AllenDynes} yields $T_\mathrm{c}= 68$ K for $E_\mathrm{F}=-50$~meV. In what follows, we analyze the possibility for conventional superconductivity in C$_3$N$_4$ in more detail.

\subsection{Coulomb interaction}
Before we proceed to a more rigorous analysis of the superconductivity in C$_3$N$_4$, we consider the problem of the electron-electron repulsion, which is another crucial factor for conventional superconductivity.

\begin{figure}[t]
\centering
\mbox{
\includegraphics[width=0.50\linewidth]{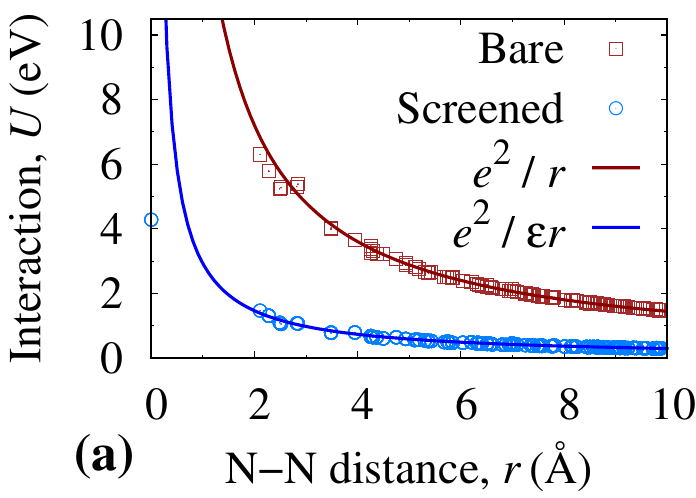}
\includegraphics[width=0.50\linewidth]{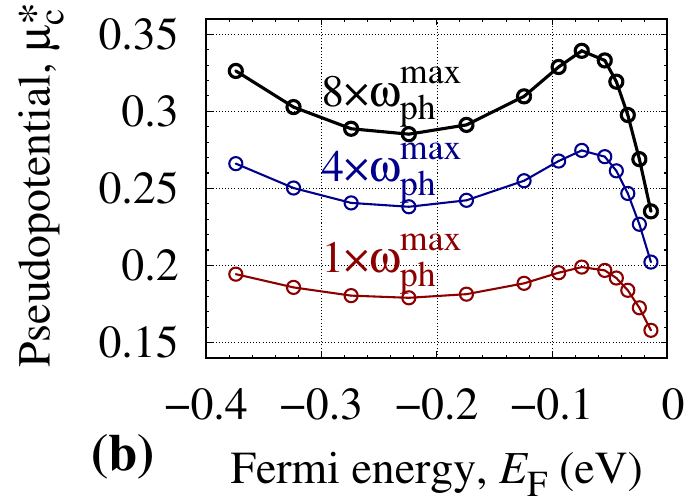}
}
\caption{
(a) Bare and screened Coulomb interaction calculated between the relevant $p_z$-like orbitals of N in C$_3$N$_4$ as a function of the interorbital distance $r$. Lines are obtained by fitting to the classical Coulomb law. (b) Coulomb pseudopotential $\mu_\mathrm{C}^*$ calculated as a function of the Fermi energy $E_\mathrm{F}$ shown for different cutoff energies of phonons: $\omega_{\mathrm{ph}}=$1$\times\omega^{\mathrm{max}}_{\mathrm{ph}}$, 4$\times\omega^{\mathrm{max}}_{\mathrm{ph}}$, and 8$\times\omega^{\mathrm{max}}_{\mathrm{ph}}$.}
\label{coulomb}
\end{figure}

The electron-electron interaction constant averaged over the Fermi surface can be written as
\begin{equation}
    \mu_\mathrm{C} = N_\mathrm{F}^{-1}\sum_{mn}\sum_{{\bf k}{\bf k}'} W_{{\bf k}-{\bf k}'} \delta(E_{n{\bf k}} - E_\mathrm{F})  \delta(E_{m{\bf k}'}- E_\mathrm{F}),
\end{equation}
where $N_\mathrm{F} = \sum_{n{\bf k}}{\delta(E_{n{\bf k}} - E_\mathrm{F})}$ is DOS at the Fermi energy.
To estimate the Coulomb interaction, we employ the random phase approximation (RPA) \cite{GrafVogl,PhysRevB.77.085122}, in which $W_{\bf q} = V_{\bf q}(1-\Pi_{\bf q}V_{\bf q})^{-1}$ with $V_{\bf q}$ being the bare (unscreened) interaction, and $\Pi_{\bf q}$ being the static polarizability.
We first consider the Coulomb interaction in C$_3$N$_4$ without doping. To this end, we calculate the bare $V(r_{ij})=\langle \phi_i \phi_i | r^{-1}_{ij} | \phi_j \phi_j \rangle$ and screened $U(r_{ij})$ Coulomb interaction between the relevant N-$p_z$ orbitals as a function of the interorbital distance $r_{ij}$, shown in Fig.~\ref{coulomb}(a). The screened interaction is calculated within RPA $U_{\bf q} = V_{\bf q} + V_{\bf q}\Pi^{(0)}_{\bf q} U_{\bf q}$ with $\Pi^{(0)}_{\bf q}$ being the static polarizability of undoped C$_3$N$_4$. Due to sufficient localization of the N orbitals, the bare interaction fits very well to the classical $V(r)=e^2/r$ law except the point $r=0$, which is irrelevant for the purpose of our study. To a good approximation, the screened interaction $U(r)$ can be obtained by the rescaling $V(r)\rightarrow V(r)/\varepsilon$, which allows us to determine the dielectric constant in undoped C$_3$N$_4$, which is found to be $\varepsilon=4.9$.
Upon the doping, there is another (metallic) contribution to the screening, which is essentially ${\bf q}$- and doping-dependent. In order to take this contribution into account, we calculate the metallic contribution to the polarizability $\Pi^{(1)}_{\bf q}$ considering the N valence states separately, within the model discussed earlier. The resulting Coulomb interaction 
can be then written as 
$W_{\bf q}=U_{\bf q}(1-\Pi^{(1)}_{\bf q}U_{\bf q})^{-1}$, where $U_{\bf q}=4\pi e^2/\varepsilon q^2.$

\begin{figure}[t]
\centering
\includegraphics[width=1.04\linewidth]{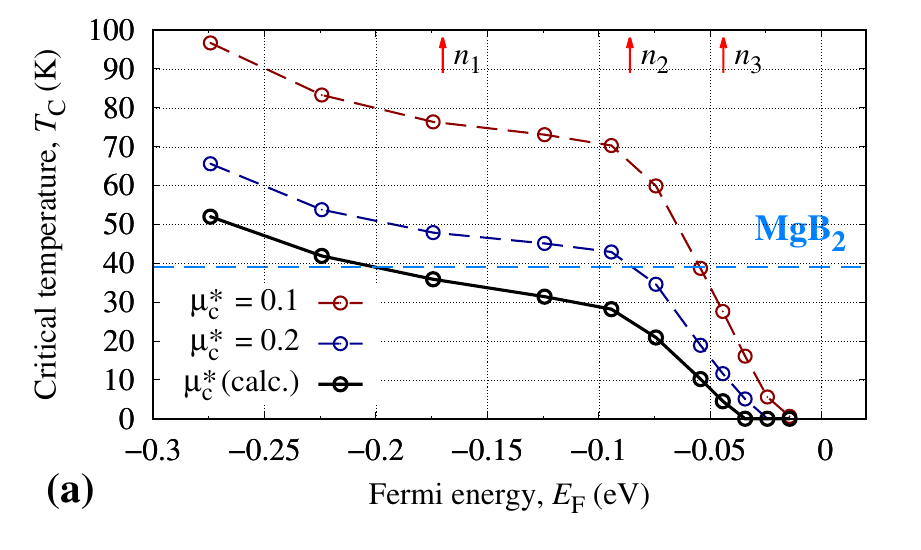}
\mbox{
\includegraphics[width=0.51\linewidth]{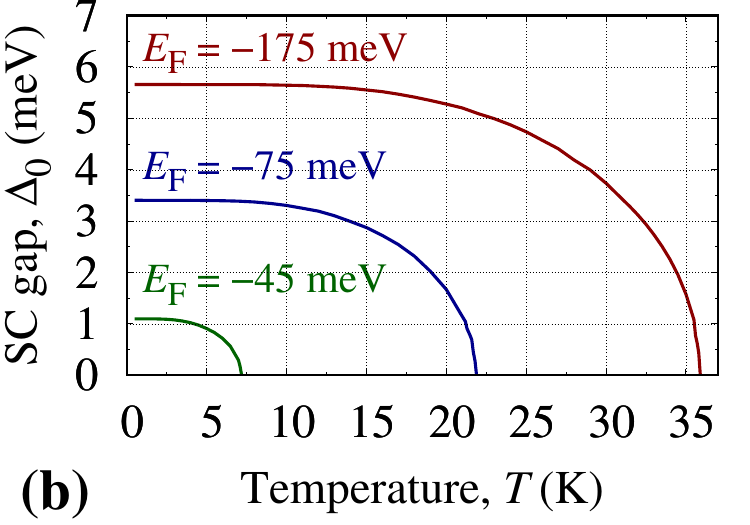}
\includegraphics[width=0.51\linewidth]{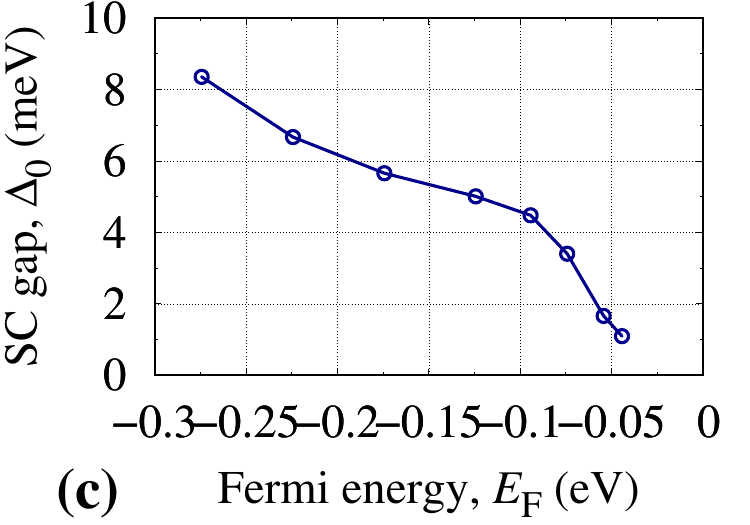}
}
\caption{
(a) Critical temperature obtained from solving the isotropic Eliashberg equations for C$_3$N$_4$ for different $E_\mathrm{F}$. The results for the Coulomb potentials $\mu_\mathrm{C}^*=0.1$ and 0.2 are shown for reference. Red arrows indicate the Fermi energies that are achievable through the doping by boron with the concentrations $n = 6$, 3, and 2 B atoms per nm$^{3}$. (b) Energy gap of the superconducting state $\Delta_0$ as a function of temperature for three different $E_\mathrm{F}$. (c) $\Delta_0$ calculated at $T=0$ K as a function of $E_\mathrm{F}$. Horizontal blue line shows $T_\mathrm{c}$ of MgB$_2$ given for reference.}
\label{tc}
\end{figure}

In the Eliashberg theory of superconductivity \cite{Marsiglio2020} it is a common practice to deal with the logarithmically corrected Coulomb interactions, which captures the fact that the interaction decays away from the Fermi surface. 
Here, we use the prescription of Marsiglio \cite{Marsiglio1989,Marsiglio1992}, who obtained the following pseudopotential for the case of a finite bandwidth $W$ and arbitrary filling,
\begin{equation}
    \mu_\mathrm{C}^* = \frac{\mu_\mathrm{C}}{1 + \frac{1}{2}\mu_\mathrm{C}\,\mathrm{ln}[E_F/\omega_{\mathrm{ph}}] + \frac{1}{2}\mu_\mathrm{C}\,\mathrm{ln}[(W-E_F)/\omega_{\mathrm{ph}}]},
    \label{mu_star}
\end{equation}
where $\omega_\mathrm{ph}$ is the energy cutoff for phonons, determined by the highest Matsubara frequency used to solve the Eliashberg equations. 
Equation~(\ref{mu_star}) is essentially different from the well-known expression for the Coulomb pseudopotential proposed originally by Tolmachev \cite{Tolmachev}, and Morel and Anderson \cite{MorelAnderson}. In the original formulation, $\mu^*$ is derived under the assumption of half filling, which is qualitatively well justified for metals. In this case, $E_F \simeq W/2$ and Eq.~(\ref{mu_star}) reduces to the well-known expression $\mu_C^* = \mu_C / (1+\mu_C\,\mathrm{ln}[E_F/\omega_{\mathrm{ph}}])$. In our case, we deal with a doped semiconductor where $E_F \ll W$, making the ``standard'' formulation inapplicable. In our calculations, we set $W=24.3$ eV, which corresponds to the width of the valence band formed by the $2s$ and $2p$ orbitals of C and N.

The resulting $\mu_\mathrm{C}^*$ is shown in Fig.~\ref{coulomb}(b) as a function of the Fermi energy for three different cutoff energies of phonons. The case $\omega_{\mathrm{ph}}=1\times\omega^{\mathrm{max}}_{\mathrm{ph}}$ corresponds to the typical choice used in conjunction with the Allen-Dynes formula for $T_c$, showing a variation between 0.15 and 0.2, depending on the Fermi energy. In the present study, in order to solve the Eliashberg equations, we use a larger cutoff, namely, $\omega_{\mathrm{ph}}=8\times\omega^{\mathrm{max}}_{\mathrm{ph}}\simeq 1.3$~eV. This choice ensures that the estimated critical temperatures are converged for each $E_F$ considered, as demonstrated in the SM \cite{SM}.

\subsection{Phonon-mediated superconductivity}
We now determine temperature dependence of the superconducting gap $\Delta_0(T)$ as well as its dependence of the Fermi energy in C$_3$N$_4$. To this end, we solve the isotropic Eliashberg equations \cite{Marsiglio2020} on the imaginary axis using the coupling parameters obtained earlier. The corresponding equations can be found in the SM \cite{SM}.  

Strictly speaking, the Migdal-Eliashberg theory has limited applicability in case of strong electron-phonon coupling. However, recent determinant quantum Monte Carlo calculations for the Holstein model demonstrate that the results quantitatively agree with the Migdal-Eliashberg theory as long as the renormalized electron-phonon coupling $\lambda \lesssim 1.7$ \cite{Kivelson2018,Kivelson2020}. This critical value is far beyond the values considered in the present study [see Fig.~\ref{phonons}(c)], suggesting that the Migdal-Eliashberg theory is adequate for moderately doped C$_3$N$_4$.

Figure~\ref{tc}(a) shows the calculated critical temperature in C$_3$N$_4$ as a function of the Fermi energy. The results are shown for the calculated $\mu_\mathrm{C}^*$ calculated for $\omega_{\mathrm{ph}}=8\times\omega^{\mathrm{max}}_{\mathrm{ph}}$ [Fig.~\ref{coulomb}(b)], as well as for the two reference values $\mu_\mathrm{C}^*=0.1$ and 0.2, representing boundaries of the typically used Coulomb pseudopotential. The dependence $T_\mathrm{c}$ versus $E_\mathrm{F}$ resembles electronic DOS with a steep increase up to $-0.1$ eV followed by a moderate growth. Already at $E_\mathrm{F} = -0.05$~meV, we obtain $T_\mathrm{c}\approx 8$ K, increasing up to $\sim 50$~K at $E_\mathrm{F} = -0.25$~eV. For $E_\mathrm{F} \lesssim -0.20$~eV, the critical temperature is found to be larger than in the prototypical conventional superconductor MgB$_2$.
Figure~\ref{tc}(b) displays a typical temperature dependence of the superconducting gap $\Delta_0$ for three representative Fermi energies. Figure~\ref{tc}(c) shows $\Delta_0(T=0)$ as a function of $E_\mathrm{F}$. The gap opens up at $E_\mathrm{F} \sim -30$~meV, which corresponds to a hole concentration of $\sim$4$\times$10$^{20}$ cm$^{-3}$. At zero temperature, we obtain the ratio $2\Delta_0/T_\mathrm{c}$ in the range 3.7--3.8, which lies in between the weak (3.53) and strong (4) coupling regimes in the BCS theory \cite{Marsiglio2001}.  

The predicted critical temperatures in doped C$_3$N$_4$ are fairly large for conventional superconductivity, and may even reach liquid nitrogen temperature at high enough doping. Up to now, we did not specify how the hole doping could be realized in C$_3$N$_4$. In analogy to diamond \cite{Abrikosov2006,Abrikosov2008}, where boron demonstrates the ability to substitute carbon atoms and induce hole doping without significant changes in the electronic structure, we propose a similar mechanism for C$_3$N$_4$, which is discussed in the next section.

\subsection{Realization of hole doping in C$_3$N$_4$}
Unlike diamond, there are two possibilities for the substitutional doping in C$_3$N$_4$, where dopants may replace C or N atoms. In order to check whether this process is energetically favorable, we calculated the formation energy of a boron dopant in C$_3$N$_4$, considering the replacement of C and N atoms individually. The formation energy of boron dopants in C$_3$N$_4$ can be calculated as \cite{Zhang1991}
\begin{equation}
    E_{\mathrm{f}} = E_{\mathrm{D}} - E_{\mathrm{P}} + \mu_{\mathrm{C/N}} - \mu_{\mathrm{B}},
    \label{eform}
\end{equation}
where $E_D$ is the energy of a defective supercell with a boron dopant, $E_{\mathrm{P}}$ is the energy of a supercell without defects. $\mu_{\mathrm{C/N}}$ is the chemical potential of a replaced C or N atom, and $\mu_{\mathrm{B}}$ is the chemical potential of the B dopant. In Eq.~(\ref{eform}), we do not explicitly consider the charge transfer term because here we only deal with a small hole doping induced by B. Indeed, the maximum possible charge transfer induced by B in the highest considered concentration (6$\times$B / nm$^3$) is 0.25 $|e|$, meaning a marginal contribution to the formation energy.

\begin{table}[tbp]
    \caption{Formation energies $E_\mathrm{f}$ (in eV) of boron dopants in C$_3$N$_4$ calculated for three different supercells, corresponding to the concentration of 2, 3, and 6 B atoms per nm$^{3}$. The values are given for the substitution of C and N atoms in the C-rich and C-poor conditions.}
\begin{ruledtabular}
\begin{tabular}{ccccc}
\label{ef-table}
& \multicolumn{2}{c}{C substitution} & \multicolumn{2}{c}{N substitution} \\
 \cline{2-3} \cline{4-5}
\textrm{Concentration}&
\textrm{C-rich }&
\textrm{C-poor }&
\textrm{C-rich }&
\textrm{C-poor } \\
\colrule
2$\times$B / nm$^{3}$ & $-2.41$   &  $-1.14$  & $+0.49$ &   $-0.46$ \\
3$\times$B / nm$^{3}$ & $-2.42$ &  $-1.15$  &  $+0.51$ &  $-0.45$ \\
6$\times$B / nm$^{3}$ & $-1.85$  &  $-0.58$  & $+0.57$ &   $-0.38$
\end{tabular}
\end{ruledtabular}
\end{table}

%In this work, we assume that B atoms can substitute either C or N atoms in C$_3$N$_4$, and calculate the formation energies of these two processes. 
To calculate the formation energies, we use three different supercells, corresponding to the concentration of 6, 3, and 2 B atoms per nm$^{3}$. Specifically, we employ 
(1$\times$2$\times$2), (2$\times$2$\times$2), and (2$\times$2$\times$3) supercells with the compositions C$_{11}$BN$_{16}$, C$_{23}$BN$_{32}$, C$_{35}$BN$_{48}$ for the C$\rightarrow$B substitution, and C$_{12}$BN$_{15}$, C$_{24}$BN$_{31}$, C$_{36}$BN$_{47}$ for the N$\rightarrow$B substitution. In each case, the atomic positions in the supercell containing a dopant were relaxed, keeping the lattice constants fixed. The presence of B dopants has a minor effect on the position of surrounding atoms. The chemical potentials $\mu_\mathrm{C}$ and $\mu_\mathrm{N}$ in Eq.~(\ref{eform}) are dependent on the formation conditions. In our calculations, we consider C-rich and C-poor conditions. In the former case, we take $\mu_{\mathrm{C}}=\mu^{\mathrm{bulk}}_{\mathrm{C}}$, where $\mu^\mathrm{bulk}_{\mathrm{C}}$ is the chemical potential of C atoms in diamond. In this case, $\mu_\mathrm{N}$ is determined from the condition 
\begin{equation}
\mu_{\mathrm{C_3N_4}} = 3\mu_{\mathrm{C}} + 4\mu_{\mathrm{N}},
\end{equation}
where $\mu_{\mathrm{C_3N_4}}$ is the energy of pristine C$_3$N$_4$ per formula unit. In the C-poor case, $\mu_\mathrm{N}=\mu^{\mathrm{bulk}}_\mathrm{N}$, where $\mu^{\mathrm{bulk}}$ is the chemical potential of N atoms in the $\alpha$ phase of crystalline nitrogen. Similarly, $\mu_\mathrm{C}$ is determined from the condition given above. In all cases, we take $\mu_{\mathrm{B}}=\mu^\mathrm{bulk}_{\mathrm{B}}$, where $\mu^\mathrm{bulk}_{\mathrm{B}}$ is the chemical of B atoms in the $\alpha$ phase of rhombohedral boron. 

The calculated formation energies are given in Table~\ref{ef-table}. We find that the substitution of C atoms by boron is considerably more favorable with the formation energies lying between $-$2.4 and $-$0.6 eV depending on dopant concentration and the formation conditions. On the contrary, the formation energy of the N substitution is found in between $-$0.5 and $+$0.6 eV, demonstrating that this type of B doping is considerably less likely.

\begin{figure}[tb]
\centering
\includegraphics[width = 0.75\linewidth]{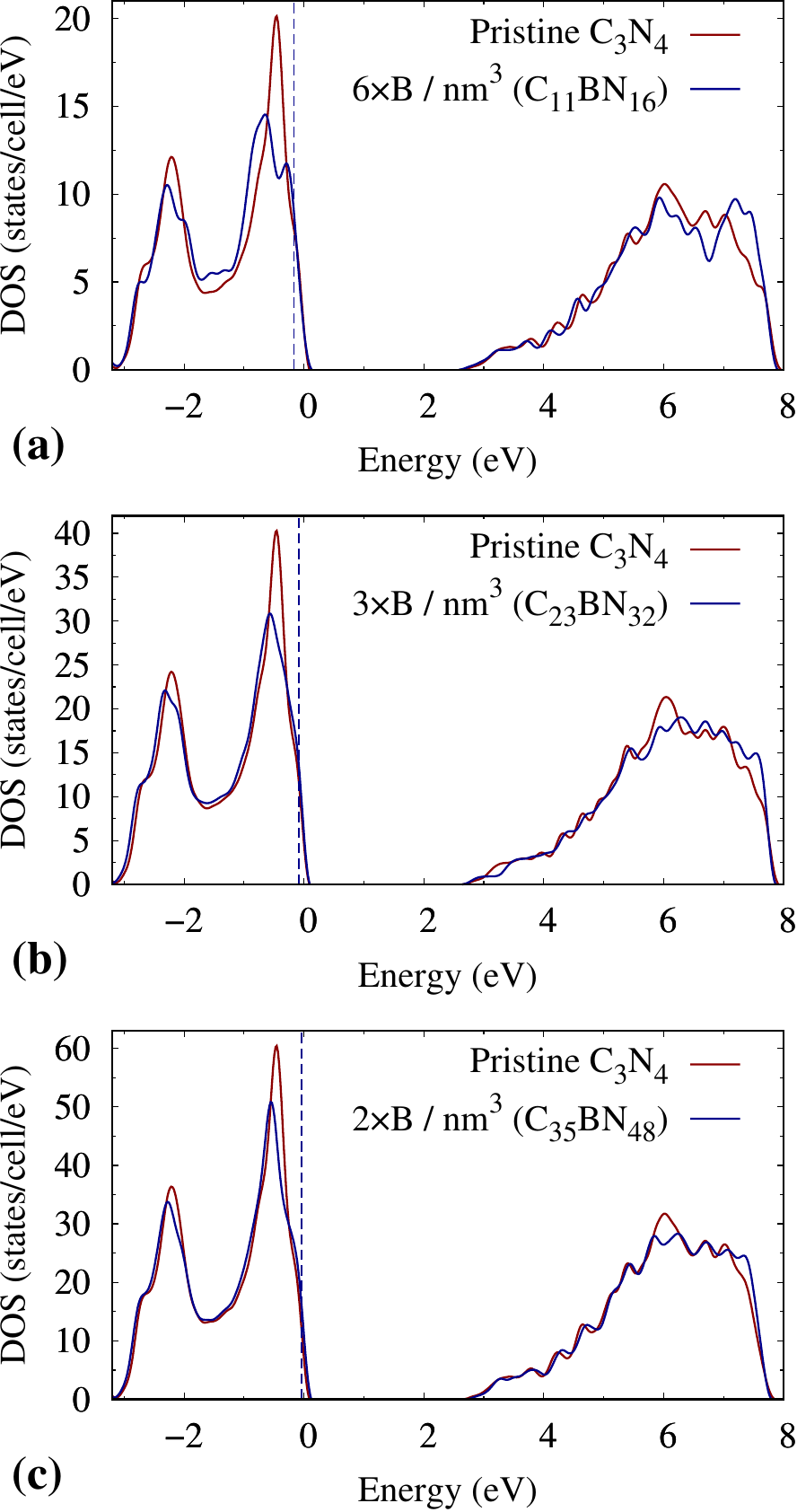}
	\caption{Electron DOS of boron-doped C$_3$N$_4$. The results are shown for three different supercells, corresponding to the concentration of 2, 3, and 6 B atoms per nm$^3$, as well as for pristine C$_3$N$_4$. In all cases, zero energy corresponds to the valence band maximum. Vertical line is the Fermi energy of the doped system. A shift of the Fermi energy from the valence band edge increases with the B concentration.}
\label{dos_defects}
\end{figure}

We now analyze the electronic structure of C$_3$N$_4$ doped by B atoms substituting C atoms. 
%To this end, we consider three different supercells with one B atom per cell, corresponding to the concentration of 6, 3, and 2 B atoms per nm$^{3}$. 
In Fig.~\ref{dos_defects}, we show the electronic DOS for the case of C$\rightarrow$B substitution corresponding to three different concentrations. DOS of pristine C$_3$N$_4$ is shown for reference in each panel.
The calculated DOS shows that the presence of boron defect does not induce any midgap states, but leads to a shift of the Fermi energy into the valence band, indicating a $p$-type doping. The shift of the Fermi energy increases with the B concentration, while the form of the electronic spectrum remains essentially unchanged. 
At the largest considered concentration of 6$\times$B / nm$^{3}$ one can see a splitting of the main peak, which takes place below the Fermi energy, which is not affecting the states in the immediate vicinity of the Fermi energy. This demonstrates that the rigid shift model of the hole doping in C$_3$N$_4$ is well justified up to the boron concentrations of 6$\times$B / nm$^{3}$,
which allows one to achieve the hole density of $\sim$10$^{22}$ cm$^{-3}$ (see Table~S1 in the SM \cite{SM} for a correspondence between the units for different B concentrations). Under these conditions, the superconducting critical temperature is estimated to be around 36 K (Fig.~\ref{tc}). At higher B dopings, a more significant modification of the electronic structure is expected, limiting the applicability of the rigid band-shift approximation.

Lastly, we comment on the dynamical stability of C$_3$N$_4$ in the presence of hole doping.
The presence of van Hove singularities and related anomalies in the electronic spectrum often causes softening of the phonon modes, developing dynamical instabilities of the atomic structures \cite{Lugovskoi2019a,Lugovskoi2019b}. In order to rule out this scenario for C$_3$N$_4$, we calculate the phonon dispersion under the hole doping induced by adding a positive charge to the system. Our calculations presented in Fig.~S1 \cite{SM} demonstrate the absence of imaginary phonon modes in C$_3$N$_4$ up to the concentrations of at least 0.4$\times$10$^{23}$ cm$^{-3}$, which suggests that the system remains dynamically stable in the relevant range of hole concentrations. Beyond this range, the dynamical stability of C$_3$N$_4$ was not examined.

%The calculations of the formation energies reveal that the substitution of C by B in C$_3$N$_4$ is considerably more energetically favorable than the substitution of N. 
%The doping of boron induces a shift of the Fermi energy toward negative energies, which increases with the B concentration. The form of the electronic spectrum remains essentially unchanged in the presence of doping. 

\section{Conclusion\label{conclusion}}
Summarizing, we performed a systematic study of the electron-phonon interaction and superconductivity in a recently discovered tetragonal phase of the carbon nitride C$_3$N$_4$ \cite{Laniel2023}. C$_3$N$_4$ features flat-band anomalies in the electronic spectrum, suggesting the development of many-body instabilities upon hole doping. Using state-of-the-art first-principles calculations, we demonstrated that hole-doped C$_3$N$_4$ exhibits strong electron-phonon coupling with the constant $\lambda \lesssim 1$, originating predominantly from the high-frequency part ($\gtrsim$100 meV) of the phonon spectrum. In turn, large $\lambda$ permits the emergence of conventional superconductivity even at moderate dopings. The critical temperature of the superconducting transition is estimated from solving the Eliashberg equations on the imaginary axis. For this purpose, we carefully estimated the Coulomb pseudopotential $\mu^*_\mathrm{C}$, which turns out to be doping-dependent, lying around 0.3 for the relevant doping regime and phonon energy cutoff. Our calculations show that $T_\mathrm{c}$ steadily increase with the hole concentration, reaching $T_\mathrm{c} \simeq 36$~K at $n_\mathrm{h}\simeq 1\times10^{22}$~cm$^{-3}$. We demonstrate that this concentration could be achieved by doping C$_3$N$_4$ with boron. At moderate concentrations up to 6$\times$B / nm$^{3}$, the electronic structure of C$_3$N$_4$ remains essentially unchanged with the exception of the Fermi energy shift, manifesting itself a $p$-type doping. At the same time, we do not exclude that even higher doping levels, i.e., higher $T_\mathrm{c}$, could be reached by other dopants.

Our findings expand the spectrum of ambient-pressure conventional superconductors with relatively high $T_\mathrm{c}$, and call for its experimental verification. We note that the presence of flat bands in C$_3$N$_4$ might trigger other many-body instabilities even at higher temperatures, potentially leading to the formation of different exotic states of matter such as charge density waves or $sp$ magnetism. This paves the way for further research in this direction. Finally, it is worth mentioning that multiple other phases of the recently discovered carbon nitrides \cite{Laniel2023} with ultra-incompressibility also represent a promising playground for further superconductivity studies.

\begin{acknowledgements}
The work was supported by the European Research Council via Synergy Grant No. 854843, FASTCORR.
I.A.A. acknowledges support as a Wallenberg Scholar (Grant No. KAW-2018.0194). I.A.A. also acknowledges support provided by the Swedish Research Council (VR), Grant No. 2019-05600, and by the Swedish Government Strategic Research Area in Materials Science on Functional Materials at Linköping University (Faculty Grant SFO-Mat-LiU No. 2009 00971).
\end{acknowledgements}

\bibliography{ref}% Produces the bibliography via BibTeX.

\maketitle

\onecolumngrid\newpage\twocolumngrid
%\clearpage
\onecolumngrid
\begin{center}
\textbf{\large Supplemental Material: Strong electron-phonon coupling and phonon-induced superconductivity in tetragonal C$_3$N$_4$ with hole doping}
\newline
\end{center}
\twocolumngrid

\setcounter{equation}{0}
\setcounter{figure}{0}
\setcounter{table}{0}
\makeatletter
\renewcommand{\theequation}{S\arabic{equation}}
\renewcommand{\thefigure}{S\arabic{figure}}
\setcounter{subsection}{0}

\subsection{Isotropic Eliashberg equations}
Below, we provide an expression for the Eliashberg equations on the imaginary axis written in the isotropic formulation \cite{Margine2013} used in these work to estimate temperature dependence of the the superconducting gap.
\begin{equation}
Z(i\omega_n) = 1 + \frac{\pi T}{\omega_n}\sum_{n'}\frac{\omega_{n'}}{\sqrt{\omega_n^2+\Delta^2(i\omega_n)}}\lambda(n-n')
\label{eeq1}
\end{equation}
\begin{equation}
    Z(i\omega_n)\Delta(i\omega_n) = \pi T\sum_{n'}\frac{\Delta(i\omega_{n'})}{\sqrt{\omega_n^2+\Delta^2(i\omega_n)}}[\lambda(n-n')-\mu_\mathrm{C}^*],
    \label{eeq2}
\end{equation}
where
\begin{equation}
    \lambda(n-n') = \int_0^{\infty} d\omega \frac{2\omega \alpha^2F(\omega)}{(\omega_n - \omega_{n'})^2 + \omega^2}.
\end{equation}
In the equations above, $i \omega_n = i(2n+1)\pi T$ is the fermionic Matsubara frequency, and the summation in Eqs.~(\ref{eeq1}) and (\ref{eeq2}) is running up to the cutoff frequency determined in our work as $\omega_\mathrm{ph} = 8\times\omega^\mathrm{max}_\mathrm{ph}$ with $\omega^\mathrm{max}_\mathrm{ph}$ being the maximum phonon frequency. The corresponding convergence tests are presented in Fig.~\ref{tc-conv}.

\begin{figure}[b]
\centering
\includegraphics[width = 0.80\linewidth]{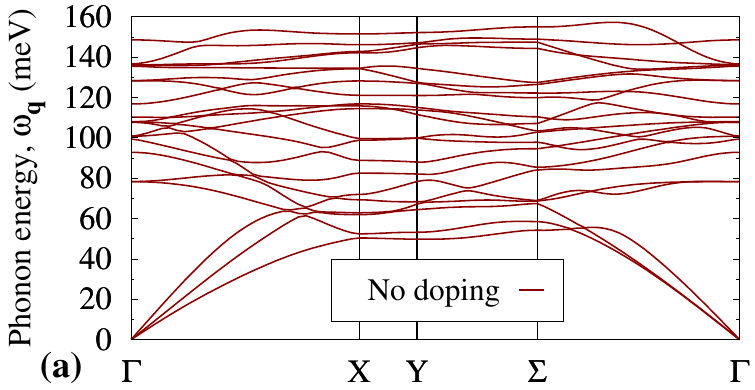}
\includegraphics[width = 0.80\linewidth]{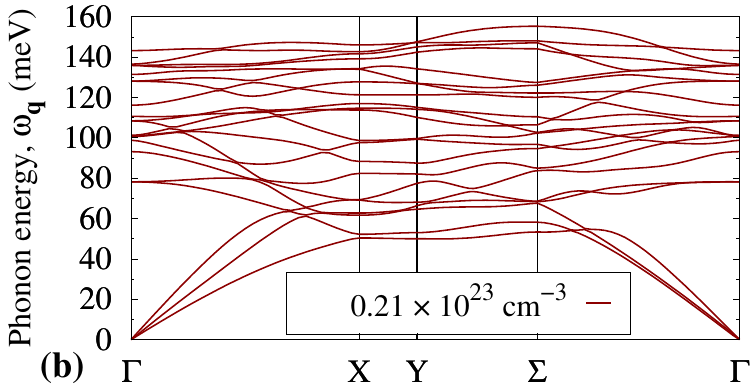}
\includegraphics[width = 0.80\linewidth]{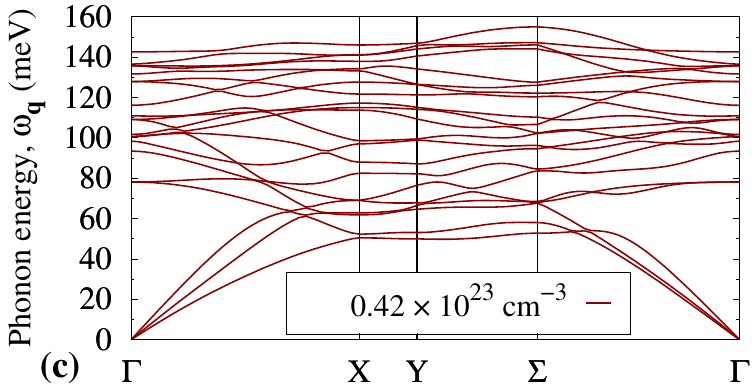}
\caption{Phonon dispersion calculated for C$_3$N$_4$ under the hole doping induced by adding a positive charge to the system. The dispersion in (a) is calculated without additional charge, whereas (b) and (c) are obtained in the presence of +0.05 and +0.1 $|e|$ per unit cell, respectively. The absence of imaginary modes indicates dynamical stability of moderately doped C$_3$N$_4$.}
\label{phonons_doping}
\end{figure}

\subsection{Dynamical stability of hole-doped C$_3$N$_4$}
Figure~\ref{phonons_doping} shows phonon dispersion calculated for C$_3$N$_4$ under the hole doping induced by adding a positive charge to the system. The dispersion in Fig.~\ref{phonons_doping}(a) is calculated without additional charge, whereas Figs.~\ref{phonons_doping}(b) and \ref{phonons_doping}(c) are obtained in the presence of +0.05 and +0.1 $|e|$ per unit cell, respectively.

\subsection{Comparison between {\sc vasp} and {\sc quantum espresso}}

\begin{figure}[tbp]
\centering
\includegraphics[width = 0.76\linewidth]{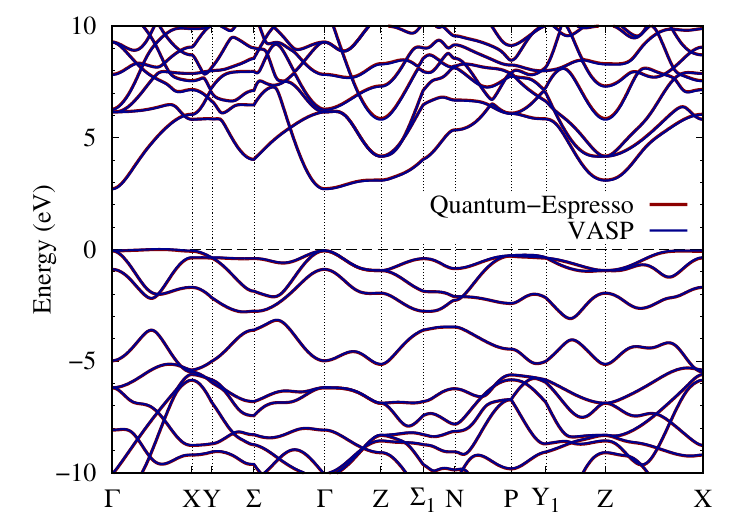}
\caption{A comparison between the band structures calculated for C$_3$N$_4$ by means of {\sc quantum espresso} and {\sc vasp} codes.}
\label{comparison}
\end{figure}

In our study, we use two different computational schemes based on which we estimate the electron-phonon coupling and the Coulomb interaction in C$_3$N$_4$. The first approach relies on the plane-wave pseudopotential method as implemented in {\sc quantum espresso}, while the second is based on the projector augmented wave method as implemented in {\sc vasp}. In order to make sure that these two approaches are equivalent in terms of the starting point, we compare the band structures calculated using these two methods.
In Fig.~\ref{comparison}, we show the calculated bands, which demonstrate that the two method produce almost identical results.

\subsection{Convergence tests}

\begin{figure}[tbp]
\centering
\mbox{
\includegraphics[width = 0.50\linewidth]{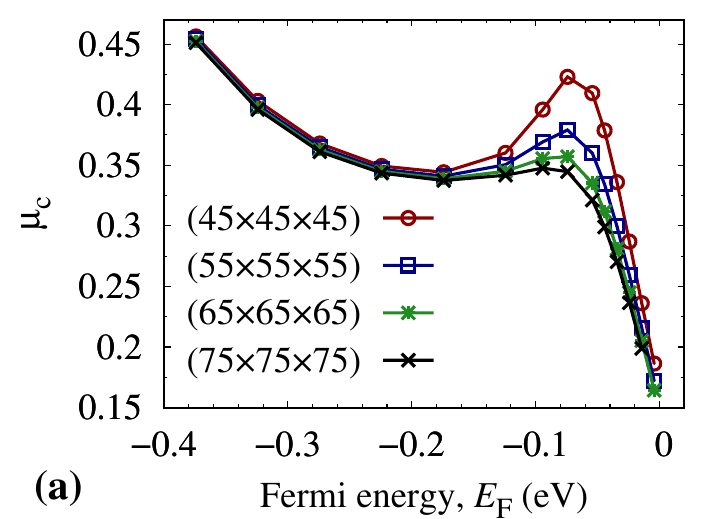}
\includegraphics[width = 0.50\linewidth]{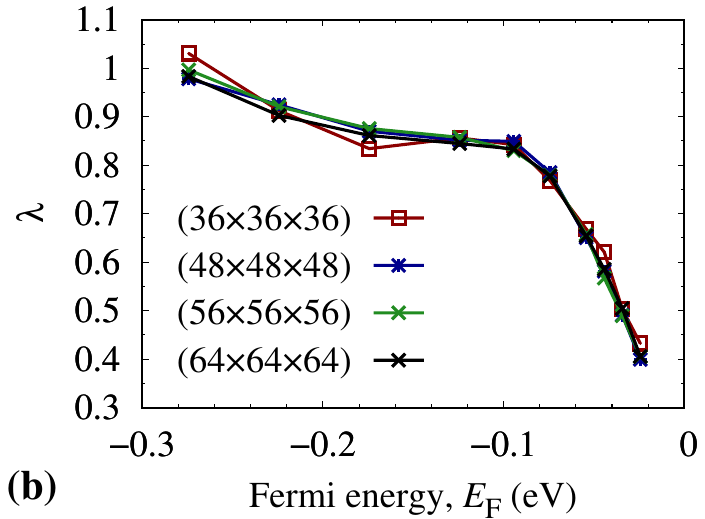}
}
\caption{Numerical convergence of the Brillouin zone integrals with respect to the ${\bf k}$- and ${\bf q}$-point meshes. (a) Coulomb interaction averaged over the Fermi surface $\mu_C$ as a function of $E_F$. (b) Electron-phonon interaction constant as a function of $E_F$. The mesh given in the legend corresponds to the ${\bf k}$ points. An equally dense ${\bf q}$-point mesh was used in the $\mu_C$ calculations. For the $\lambda$ calculations, a halved ${\bf q}$-mesh density was used. A Gaussian smearing of 0.01 eV was used in all cases.}
\label{convergence}
\end{figure}

Figures \ref{convergence}(a) and \ref{convergence}(b) show numerical convergence of the averaged Coulomb interaction $\mu_C$ and the electron-phonon coupling constant $\lambda$ with respect to the density of ${\bf k}$ and ${\bf q}$ meshes. The results presented in the main text for the Coulomb pseudopotential were obtained using (75$\times$75$\times$75) ${\bf k}$-point and ${\bf q}$-point meshes. The electron-phonon coupling was presented for a (64$\times$64$\times$64) ${\bf k}$-point and (32$\times$32$\times$32) ${\bf q}$-point meshes.

\begin{figure}[tbp]
\centering
\includegraphics[width = 0.60\linewidth]{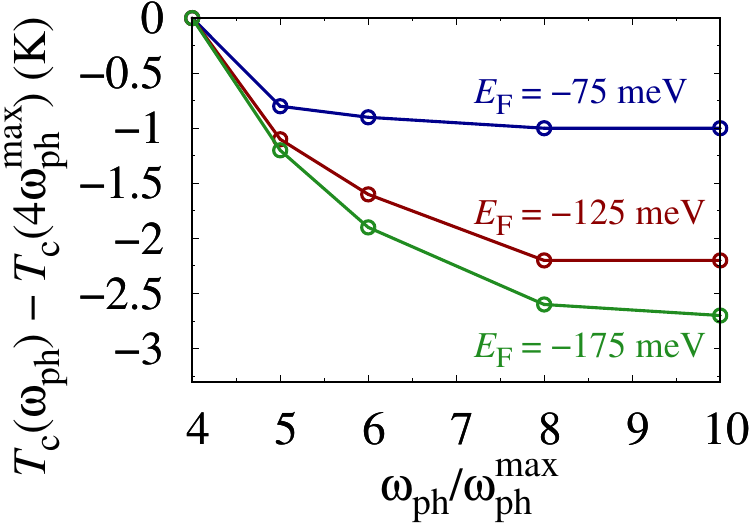}
\caption{Dependence of the critical temperature $T_c$ on the cutoff energy of phonons in the Eliashberg equations $\omega_\mathrm{ph}$ for three different Fermi energies. The $T_c$ values are given relative to the $T_c$ calculated using $\omega_{\mathrm{ph}} / \omega^{\mathrm{max}}_{\mathrm{ph}} = 4$. The results presented in the main text are obtained for $\omega_{\mathrm{ph}} / \omega^{\mathrm{max}}_{\mathrm{ph}} = 8$, which ensures numerical accuracy of $T_c$ to within 1~K.}
\label{tc-conv}
\end{figure}

Fig.~\ref{tc-conv} shows the dependence of the critical temperature on the cutoff energy of phonons in the Eliashberg equations for three different Fermi energies. The cutoff energy $\omega_{\mathrm{ph}} = 8\times\omega^{\mathrm{max}}_{\mathrm{ph}}$ used to obtain the results presented in the main text is sufficient to reach the numerical accuracy of $T_c$ to within 1~K for each $E_F$ considered.

\subsection{Fermi surfaces}
\begin{figure}[t]
\centering
\includegraphics[width = 0.97\linewidth]{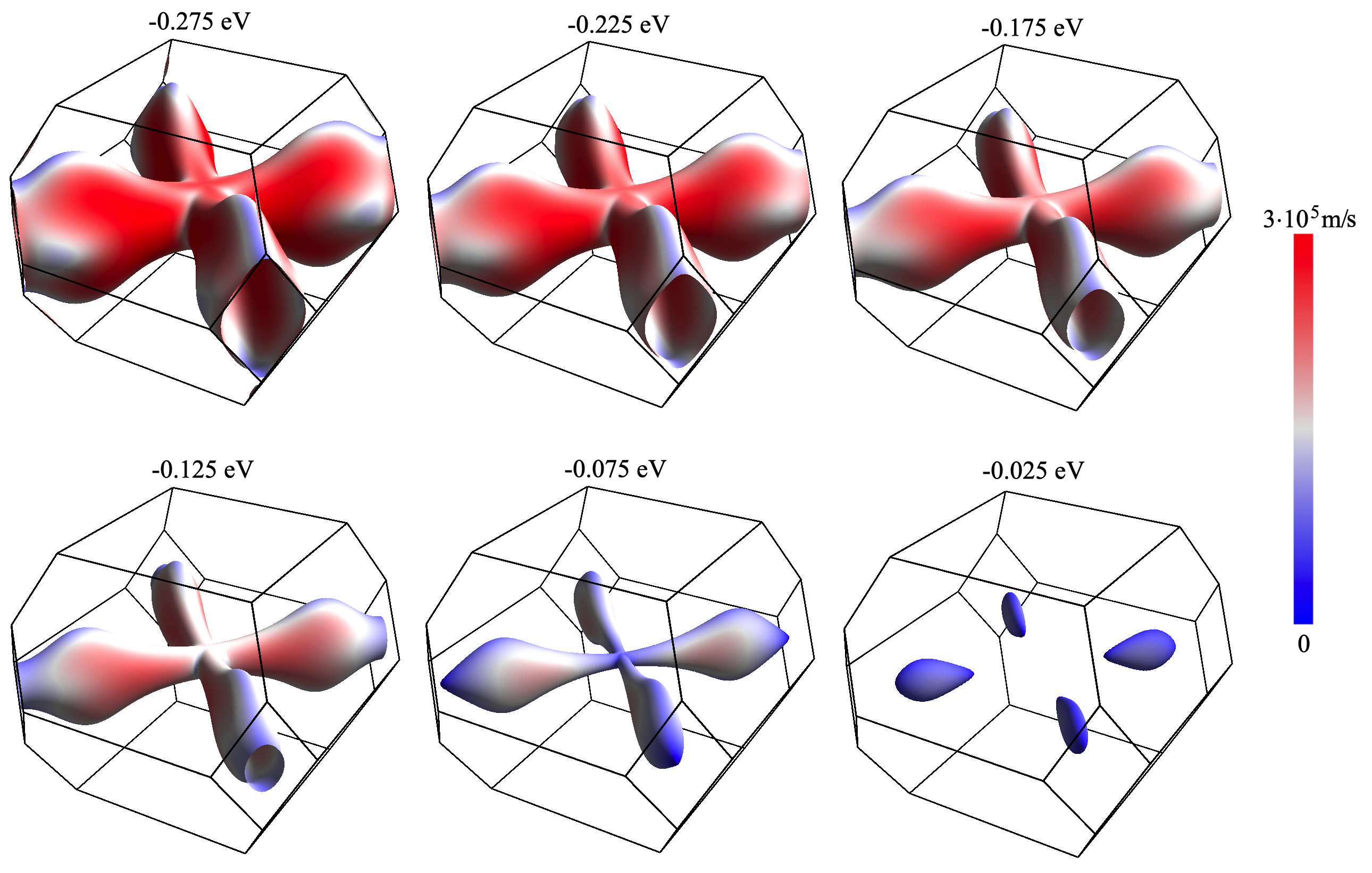}
	\caption{The Fermi surfaces of C$_3$N$_4$ calculated for a number of relevant Fermi energies, corresponding to the doping levels considered in this work. Color shows absolute values of the relative Fermi velocities, where red (blue) corresponds to the maximum (minimum) value. The figure was plotted using the {\sc fermisurfer} software \cite{KAWAMURA2019197}.}
\label{FS}
\end{figure}

Fig.~\ref{FS} shows the Fermi surfaces of C$_3$N$_4$ calculated for different Fermi energies, corresponding to the doping levels considered in this study. 

\subsection{Unit conversion}
In Table \ref{conversion}, we provide a correspondence between the units of chemical doping of boron, Fermi energy, and hole concentration in C$_3$N$_4$ for three doping levels considered in the main text.\\
\newline

\begin{table}[b]
    \caption{Conversion between the units of chemical doping of boron, Fermi energy, and  hole concentration in C$_3$N$_4$ for three doping levels considered in the main text.}
\begin{ruledtabular}
\begin{tabular}{ccccc}
\label{conversion}
	\textrm{Doping level}&
	\textrm{$n_1$}&
	\textrm{$n_2$}&
	\textrm{$n_3$} \\
\colrule
Chemical doping (B / nm$^{3}$)              &    6     &   3       &  2      \\
Fermi energy (meV)                          & $-$170   &  $-$85    &  $-$45  \\
Hole density ($\times$ 10$^{20}$ cm$^{-3}$) &   62     &  23       &  8 
\end{tabular}
\end{ruledtabular}
\end{table}

\end{document}